\documentclass[twocolumn]{24onr_art}
\usepackage{graphicx}

\pdfoutput = 1

\newcommand\etal{\mbox{\textit{et al.}}}

\newcommand\Rey{\mbox{\textit{Re}}}  
\newcommand\Fro{\mbox{\textit{Fr}}}  

\def\curve{\mbox{(~---------------~)}}
\def\curdasha{\mbox{(~------~--~------~)}}

\def\curud{\mbox{(~-~-~-~-~-~-~-~-~-~)}}

\def\curuudd{\mbox{(~~--~~--~~--~~--~~--~~)}}
\def\curuuudd{\mbox{(~~---~~---~~---~~---~~)}}

\newcommand\enetke{\ensuremath{\int_V dV \widetilde{\overline{u}}_i\widetilde{\overline{u}}_i}}
\newcommand\enemean{\ensuremath{\int_V dV \left< \overline{u}_i \right> \left<\overline{u}_i \right>}}
\newcommand\enepot{\ensuremath{\frac{1}{\Fro^2}\int_0^t d\tau \int_V dV \widetilde{\overline{\rho}}\,\widetilde{\overline{u}}_3}}

\begin{document}

%
%
%
%
%
%
%
%
\title{Numerical Simulation of Wakes in a Weakly Stratified Fluid}
%
%
%
\author{James W. Rottman$^1$, Douglas G.\ Dommermuth$^1$,
George E.\ Innis$^1$, \\ Thomas T.\ O'Shea$^1$ and Evgeny Novikov$^2$}
\affiliation{($^1$Science Applications International Corporation, USA, \\
             $^2$University of California, San Diego, USA)}

\maketitle
%
%
\section{ABSTRACT}

This paper describes some preliminary numerical studies using large eddy simulation of full-scale submarine wakes. Submarine wakes are a combination of the wake generated by a smooth slender body and a number of superimposed vortex pairs generated by various control surfaces and other body appendages. For this preliminary study, we attempt to gain some insight into the behavior of full-scale submarine wakes by computing separately the evolution the self-propelled wake of a slender body and the motion of a single vortex pair in both a non-stratified and a stratified environment. An important aspect of the simulations is the use of an iterative procedure to relax the initial turbulence field so that turbulent production and dissipation are in balance.

\section{INTRODUCTION}

The prediction and evolution of the full-scale wakes of submarines has been an objective of naval hydrodynamics for many years.  Full-scale submarine wakes consist of the wake behind a slender body on which is superimposed a number of vortex pairs generated by various control surfaces. Coherent vortex systems are major components of the wakes of submarines.  For this preliminary study, we attempt to gain some insight into the behavior of full-scale submarine wakes by using LES to compute separately the evolution of the self-propelled wake of a slender body and the motion of a single vortex pair in a stratified and turbulent environment. An important aspect of the simulations is the use of an iterative procedure to relax the initial turbulence field so that turbulent production and dissipation are in balance.

In the first section we discuss the studies of the wake of a slender, self-propelled body without any superimposed vortex pairs. The objective of this research is to be able to simulate accurately a full-scale wake of a self-propelled body using initial conditions on a cross plane a short distance downstream of the submarine. The initial conditions are obtained from laboratory experiments. In the second section we describe the simulations of a control surface vortex pair in both non-stratified and stably stratified and turbulent environments.  The ultimate objective of this study is to develop a subgrid scale turbulence model to accurately represent the turbulence in the vortex cores and to be able to predict the turbulent entrainment into and detrainment out of the recirculation region of the pair and therefore to be able to accurately compute the motion of the pair at full-scale.

\section{WAKE STUDIES} \label{sec:wake_studies}

Laboratory experiments, in the absence of any background shear, have shown that the far wake behind a self-propelled body consists of very slowly evolving patches of vertical vorticity, of alternating sign, whose horizontal extent is much greater than their vertical extent. These patches are often referred to by the descriptive name of ``pancake eddies''. 

The governing parameters for this flow are the Reynolds number $\Rey = U D / \nu$ and the Froude number $\Fro = U / (N D)$, where $U$ is the speed of the body, $D$ is the diameter of the body, $\nu$ is the kinematic viscosity of the fluid, and $N=[-g/\rho_o \, \partial \varrho/\partial x_3]^{1/2}$ is the buoyancy (or Brunt-V\"{a}is\"{a}l\"{a}) frequency in which $g$ is the acceleration due to gravity, $\rho_o$ is the mean density, and $\partial \varrho/\partial x_3$ is the vertical derivative of the background mean density (assumed to be a constant).

In this paper we describe two numerical experiments: the first is an attempt to simulate the formation of eddies in late wake due to a turbulent flow for the case with $\Fro = 2.0$ and $\Rey = 10^5$.  The second is similar to the first except that the fluid is non-stratified ($\Fro = \infty$).

The numerical method is large eddy simulation (LES). The flow is initialized with a mean wake flow that includes swirl, which is based on experimental measurements of the wake behind a propeller-driven slender body, with a superimposed homogeneous turbulent flow field. A relaxation procedure is introduced to establish the proper balance between production and dissipation before the calculation is begun. There are no coherent structures imposed on the turbulence other than the gross characteristics of the mean wake flow.  The results are compared with that of a drag wake that is described in a companion paper, Dommermuth, et al.  (2002), and it is found that the two types of wakes evolve quite differently. The mean swirl generates internal waves. Once the mean swirl radiates internal waves, the turbulence production terms are considerably reduced, especially the shear stresses.   As a result, the wake persists much longer than a drag wake.   Our results show that coherent vortices appear in the late wake even though the flow is initialized without any coherent structures.

\subsection{Problem formulation}

A schematic drawing of the flow under consideration is shown in Figure~\ref{schematic} in a reference frame in which the body generating the wake is at rest. This figure serves to define much of the nomenclature used here.  The wake is considered to be statistically stationary.  Since the entire wake is too long to compute as a whole, we make the approximation that the flow can be computed within a rectangular box, with axial dimensions much smaller than the total length of the wake, that moves with the mean flow speed $U$.  Within this box the flow is computed by solving numerically the governing equations for an incompressible Boussinesq fluid using large eddy simulation (LES).

\subsubsection{Large eddy formulation of the Boussinesq equations} \label{sec:LES}

\begin{figure}
   \begin{center}
     \includegraphics[width=\linewidth]{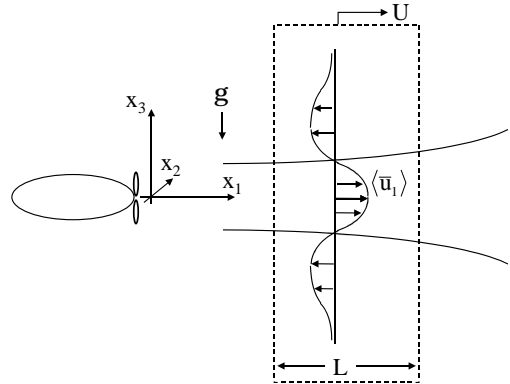}
   \end{center}
   \caption{A schematic diagram of the self-propelled body and the coordinate system in a reference frame in which the body is at rest.  The dashed box denotes the slab of fluid that is modelled using LES.  $\langle \overline{u}_1 \rangle$ is the wake deficit.}
   \label{schematic}
\end{figure}

We will assume that the fluid is incompressible and weakly stratified.  A large eddy approximation is invoked whereby the large-scale features of the flow are resolved and the small scales are modelled.  Let $\overline{u}_i$ denote the filtered three-dimensional velocity field as a function of space $x_i$ ($i = 1, 2, 3$) and time $t$. Here, the overbar denotes spatial filtering in a large eddy sense.  The origin of the coordinate system is at the centroid of the body, as shown in Figure~\ref{schematic}.  $x_1$ is positive downstream, $x_2$ is transverse to the track of the body, and $x_3$ is positive upward.  The length and velocity scales of the flow are respectively normalized by the diameter of the body ($D$) and the free-stream velocity ($U$).

The numerical method is large eddy simulation (LES). The flow is initialized with a mean wake flow with a superimposed homogeneous turbulent flow field. A relaxation procedure is introduced to allow the turbulence field to achieve a balance between production and dissipation before the calculation is begun, as is described in more detail in Dommermuth, et al. (2002). There is no coherent structure imposed on the turbulence other than the gross characteristics of the mean wake flow.

A mixed model (Bardina, {\it et al.}, 1984) is used to model the subgrid scale stress tensor. The similarity portion of the mixed model provides an accurate model of the turbulent stresses, whereas the Smagorinsky portion provides dissipation. The corresponding mixed model for the residual density flux combines a similarity model with an eddy diffusivity model.

Following Orszag \& Pao (1974), a Galilean approximation is used to relate the spatial development of the wake to the temporal evolution of the LES.   In normalized variables, this approximation results in the relation $x_1=t$, where $x_1$ is the distance downstream of the body in the wake and $t$ is the corresponding time in the LES. Based on this Galilean approximation, we further assume that

\begin{eqnarray}
\widehat{\phi}= \frac{1}{T} \int^T_0 dt \phi(t,x=X_o)
\Longleftrightarrow \nonumber \\ \frac{1}{L} \int^L_0 dx \phi(t=T_o,x) = \left<
\phi \right> \;\; ,
\end{eqnarray}

\noindent where $\phi$ is a physical quantity, hat accents denote time averaging, angle brackets denote spatial averaging, $L$ is the length of the LES computational domain in the axial direction (see Figure~\ref{schematic}), $T$ is the duration of time averaging, and $X_o$ and $T_o$ are positions in space and time where the wake of the body and the LES correspond.  A tilde accent denotes the turbulent fluctuations, which are defined as $\widetilde{\phi} = \phi - \left< \phi \right>$. As shown in Figure~\ref{schematic}, the wake of the body is modelled as a slab of fluid.

\subsubsection{Initialization}

The initial velocity field is decomposed into a mean disturbance and a fluctuating disturbance.  The magnitude and distribution of the mean and fluctuating components are specified based on the measurements of Lin \& Pao (1973, 1974a,b) and Lin, Veenhuizen \& Liu (1976). The mean axial velocity is specified as

\begin{eqnarray}
\langle \overline{u}_1 \rangle = a_o \left( 1 - \frac{r^2}{2
r_o^2} \right) \exp\left(-\frac{r^2}{2 r_o^2} \right) \;\; ,
\end{eqnarray}

\noindent where $a_o$ is the amplitude of the mean wake deficit normalized by the free-stream velocity, and $r_o$ is the initial characteristic radius of the wake.  The cross-stream components, $\langle \overline{u}_2 \rangle$ and $\langle \overline{u}_3 \rangle$ are determined from the mean axial vorticity, which is specified

\begin{eqnarray}
\langle \overline{\Omega}_1 \rangle = a_\omega \left( 1 -
\frac{r^2}{2 r_\omega^2} \right) \exp\left(-\frac{r^2}{2
r_\omega^2} \right) \;\; ,
\end{eqnarray}

\noindent where $a_\omega$ is the amplitude of the mean axial vorticity and $r_\omega$ is the characteristic radius of the axial vorticity field.  The propeller swirl consists of an annulus of vorticity of one sign from the tip vortices surrounding a region of opposite sign vorticity from the root vortices, with a net vorticity of zero. Interestingly, in another numerical experiment (not shown) we have found that the swirl is necessary if the axial velocity is to retain its self-similar shape. Without the swirl, the axial velocity rapidly disintegrates into small regions of positive and negative flow, and the rate at which they cancel each other out is more rapid than with swirl present.

The initial rms velocity fluctuations are also approximated using Gaussian distributions.

\begin{eqnarray}
\sqrt{\langle \widetilde{\overline{u}}_i
\widetilde{\overline{u}}_i \rangle} = a_i \exp\left(-\frac{r^2}{2
r_i^2} \right) \;\;
\end{eqnarray}

\noindent where $a_i$ are the initial amplitudes of the rms velocity fluctuations and $r_i$ are the initial characteristic radii.  The fluctuating velocity field is constructed from a realization of fully-developed homogeneous turbulence that is projected onto the rms velocity distribution, as is described in more detail in \cite{dommermuth97}.  The rms fluctuations are initially uncorrelated and the turbulent shear stresses are zero. As discussed later, in the {\it Wake relaxation} subsection, an iterative procedure is used to relax the wake until the production of turbulent kinetic energy is balanced by dissipation. We assume that the mean and fluctuating portions of the density disturbance are initially zero.

\subsubsection{Numerical algorithm}

The governing equations are discretized using second-order finite-differences.  A fully-staggered grid is used in the numerical simulations.  Periodic boundary conditions are used along the sides of the computational domain, and free-slip boundary conditions are imposed at the top and bottom.  A third-order Runge-Kutta scheme is used to integrate the equations with respect to time.   The numerical algorithms have been implemented using high-performance fortran (PGHPF) on a CRAY T3E. Additional details and convergence studies of a similar numerical algorithm are described in Dommermuth, \etal (1997).

\begin{figure}
   \begin{center}
      \includegraphics[width=\linewidth]{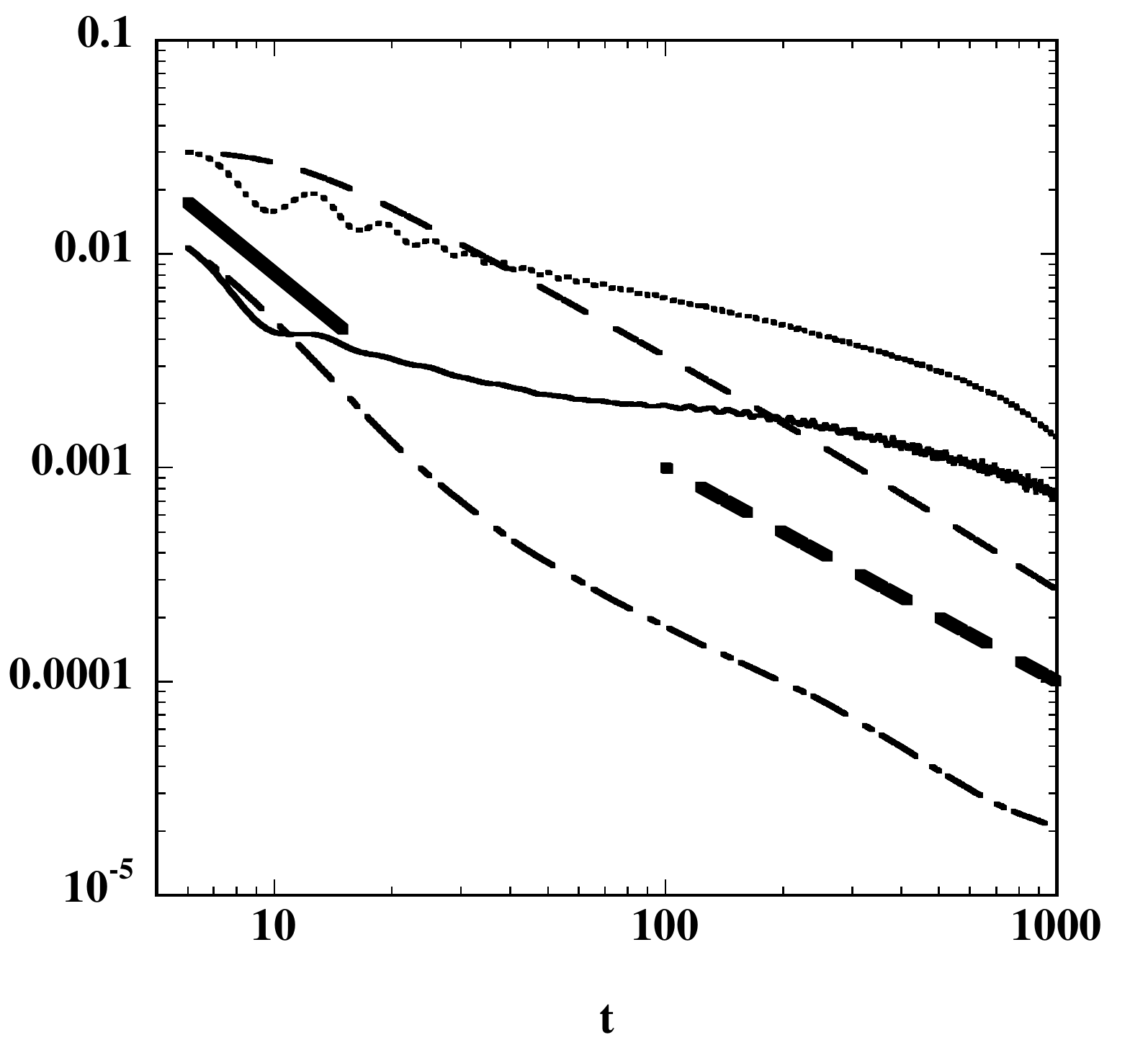}
   \end{center}
\caption{The kinetic energy for $7 < t \le 1000$ for $\Rey = 10^5$: \curuuudd, fluctuations for $\Fro = \infty,~ $\curud, fluctuations for $\Fro = 2.0$,~\curdasha, mean flow for $\Fro = \infty$,~and~\curve, mean flow for $\Fro = 2.0$. The bold solid line represents axial similarity behavior, $t^{-3/2}$, and the bold dashed line represents swirl similarity behavior, $t^{-1}$. The energy is normalized by $U^2D^3$.}
   \label{energy}
\end{figure}

\subsection{Results } \label{sec:results}

For the stratified simulation, the Reynolds number is $\Rey=10^5$ and the Froude number is $\Fro = 2.0$. The initial mean disturbance and the rms fluctuations are based on least-squares fits of laboratory measurements of a cross section of the wake that is seven diameters downstream.  For the mean axial velocity, $a_o = 0.10$ and $r_o = 0.25$ and for the mean axial vorticity $a_\Omega = 0.80$ and $r_\Omega = 0.20$.  For the fluctuating portion of the flow, pre-relaxation, $a_i = 0.40$ and $r_1 = 0.07$ and $r_2 = r_3 = 0.05$.  We choose a computational domain that is 12$D$ long, 24$D$ wide, and 12$D$ deep.  The horizontal and vertical dimensions of the computational domain are as large as computer resources allow in order to accommodate the propagation of internal waves and the spreading of the wake, which in the stratified case spreads more in the horizontal than in the vertical.  In any case, the horizontal and vertical dimensions of our computational domain are larger in most cases than the comparable tank sizes used in laboratory experiments. Convergence is established using two different grid resolutions corresponding to coarse ($128\times256\times129$ grid points) and fine ($256\times512\times257$ grid points) simulations.  The fine-grid results are presented here.  For the non-stratified case, we use the same dimensional parameter values as for the stratified case, except that $N = 0~$rad/sec. Therefore, all the nondimensional parameters are the same except that $\Fro = \infty$.

\subsubsection{Wake relaxation} \label{sec:wake}

A relaxation procedure is used to establish the proper balance between production of turbulent kinetic energy and dissipation at the beginning of the calculation.  During the relaxation procedure, the mean portion of the flow is held fixed.   The total turbulent kinetic energy is also held fixed, but the spatial distribution of the turbulent fluctuations is free to vary.   Once the turbulent production and dissipation are in balance, the relaxation procedure is turned off and the numerical simulation is initiated.  This is the same relaxation procedure described in Dommermuth, \etal (2002) and is similar to the procedure used by Orszag \& Pao (1974) in their numerical simulations of a self-propelled body.

\subsubsection{Similarity}

In Figure~\ref{energy} the kinetic energy in the mean portion of the flow integrated over the volume of fluid (\enemean) and the turbulent kinetic energy (\enetke) are plotted versus time for both the stratified and non-stratified cases for $\Rey=10^5$.

The decay of the mean and turbulent kinetic energy in the self-propelled body wake proceeds somewhat differently than for the towed body, which is described in Dommermuth, et al. (2001). Figure~\ref{energy} shows the decay of the mean and fluctuating kinetic energy for both the stratified and non-stratified wakes. Initially, for the non-stratified case, the energy in the thrust and drag wakes dominates and the energy decays as $x_1^{-3/2}$. Eventually, however, the swirl energy (which is much smaller initially) dominates and the entire wake approaches the swirl decay rate of $x_1^{-1}$.  For this particular simulation, the non-equilibrium region is much greater than it is for a drag wake. The mean and turbulent energies for the stratified simulation decay much less rapidly than the non-stratified simulation.

For the stratified simulation, self-similarity is not established until the very late wake.  We believe that once the effects of stratification disrupt the turbulent production mechanism, the wake persists significantly longer than it would in a non-stratified fluid.  This effect is more significant for momentumless wakes than it is for drag wakes because momentumless wakes in the absence of stratification decay more rapidly than drag wakes.

\begin{figure}
   \begin{center}
      \includegraphics[width=\linewidth]{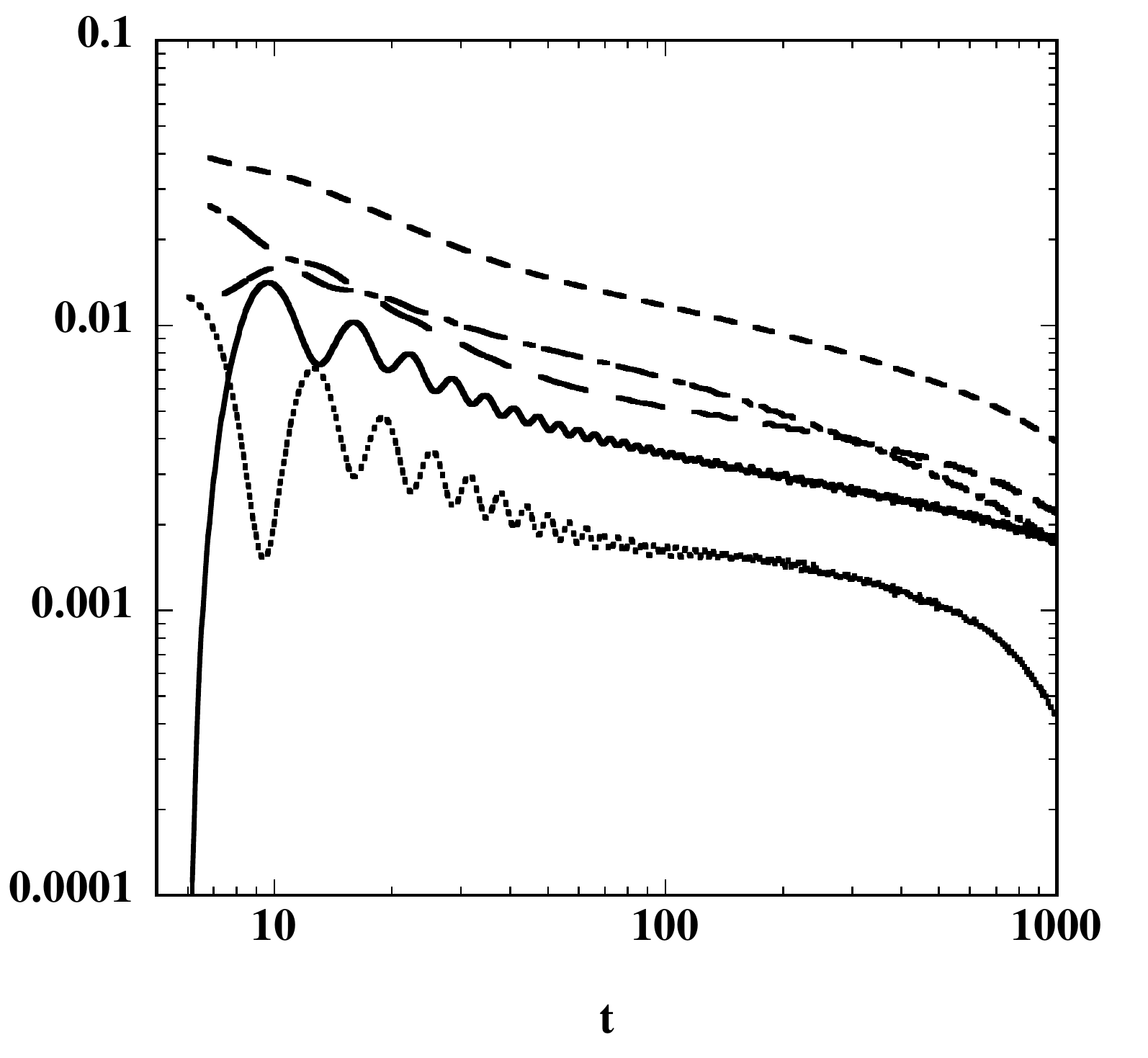}
   \end{center}
\caption{The total potential and kinetic energies for $7 < t \le 1000$ for $\Rey = 10^5$.  The results for stratified fluids ($\Fro=2.0$) are labeled: \curve, potential energy; \curud, vertical turbulent kinetic energy; \curuuudd, potential energy plus vertical turbulent kinetic energy, \curdasha, horizontal turbulent kinetic energy, and \curuudd, total energy. The energy is normalized by $U^2D^3$.}
   \label{energy_total}
\end{figure}

Energy is redistributed between the kinetic energy and the potential energy and also between the mean and the fluctuating portions of the flow.  Figure~\ref{energy_total} shows the total energy ($E$), which includes the turbulent kinetic energy, the kinetic energy in the mean portion of the flow, and the potential energy (\enepot).  Figure~\ref{energy_total} also shows the total energy in the fluctuating portion of the flow ($\widetilde{E}$), which includes the turbulent kinetic energy and the potential energy.   The stratified and non-stratified fluids establish self-similarity at the same rate for both $E$ and $\widetilde{E}$. The results show a tendency in the far wake for the energy in the stratified fluid to be higher than the non-stratified fluid. This effect may be attributed to the generation of internal waves, which do not decay as rapidly as turbulence.

\subsubsection{The formation of pancake eddies}

\begin{figure*}
   \begin{center}
     (b)~\includegraphics[width=0.55\linewidth]{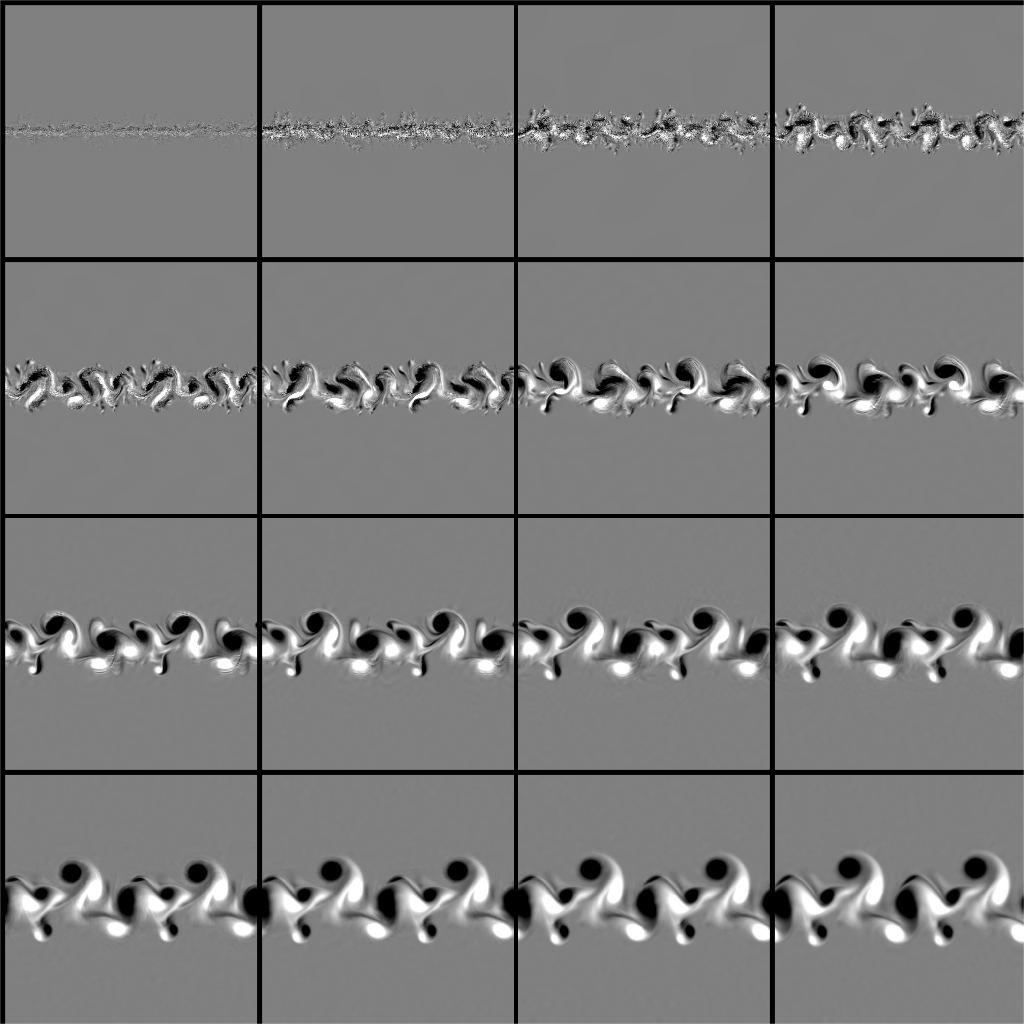}
   \end{center}
   \caption{A time history of the vertical component of vorticity    $\omega_z$ on the horizontal plane through the center of the    wake for $Re = 10^5$:(a) $\Fro=\infty$ and (b) $\Fro=2.0$. In    these grey-scale figures, white represents positive vorticity    with magnitude $\omega_z = 4$ and black negative vorticity    with $\omega_z = -4$.  Each frame has the dimensions $24 D$ in    both the cross-stream and upstream (to the left) directions.    Note that the flow along the streamwise direction ($x_1$) has    been periodically extended. The centers of each frame are    located at (from left to right and top to bottom) $t\approx$ 6,    22, 54, 86, 118, 166, 230, 294, 358, 422, 518, 614, 710, 806,    902, and 998.}
   \label{wake}
\end{figure*}

Figure~\ref{wake} show time series of the vertical component of vorticity in the horizontal plane through the wake centerline ($x_3=0$).   Part (a) illustrates the results for the non-stratified fluid, and part (b) shows the corresponding results for a stratified fluid.  In this grey-scale figure, white represents positive vorticity with magnitude $\omega_z = 4$ and black negative vorticity  with $\omega_z = -4$.  Each frame has the dimensions $24 D$ in both the cross-stream and upstream (to the left) directions. Note that the flow along the streamwise direction ($x_1$) has been periodically extended. The centers of each frame are located at (from left to right and top to bottom) $t\approx$ 6, 22, 54, 86, 118, 166, 230, 294, 358, 422, 518, 614, 710, 806, 902, and 998. Each frame is scaled by the distance downstream from the initial plane ( $t=6$ ), which is the expected similarity behavior.

For the stratified cases, coherent structures, in the form of nearly circular vortex patches begin to appear at $t \approx 100$. This corresponds to about the time the mean kinetic energy begins to decay at the self-similar rate.  Note that instabilities are evident almost immediately.  Further downstream, the size of these patches of vorticity grow and the number of patches in a frame very slowly decrease.

Over the duration of the simulation, the small-scale features that are observed in the bulges of the non-stratified simulation appear to merge to form the large-scale structures that are observed toward the end of the stratified simulation.

\begin{figure*}
   \begin{center}
     (a)~\includegraphics[width=0.55\linewidth]{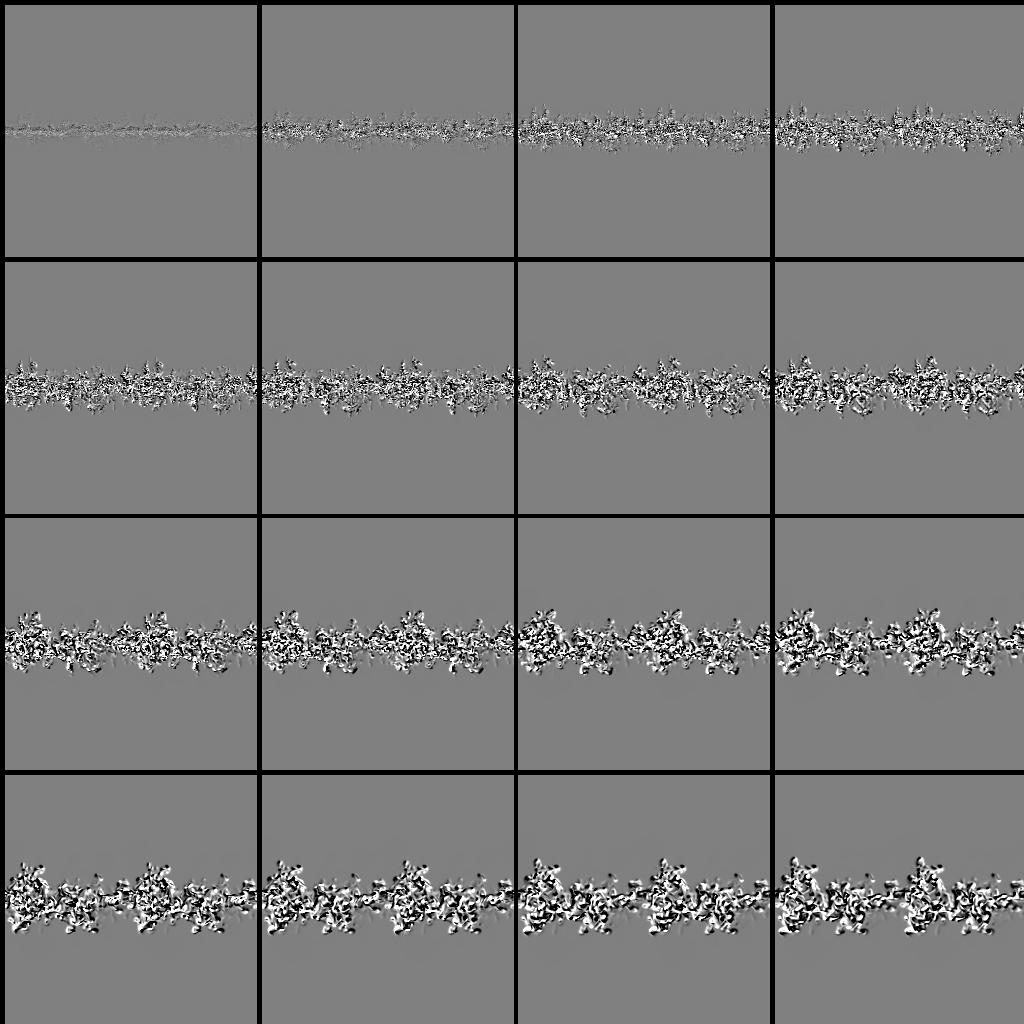}
   \end{center}
\end{figure*}

\section{VORTEX STUDIES}

Coherent vortex pairs are major components of the wakes of submarines.  Recently, comparisons with laboratory and field observations (Delisi and Greene, 1990; Delisi, \etal, 1996; Delisi, 1998) indicate that numerical models fail to accurately predict vortex pair migration in stratified environments.  The reasons for this failure are not fully understood.  A possible reason for this failure is that the turbulence models employed in the numerical simulations produce inaccurate predictions of the entrainment and detrainment rates of the recirculating region of the vortex pair.  Large-eddy-simulation (LES) codes, developed for modelling aircraft trailing vortex pairs in the atmosphere (Gerz and Ehret, 1997; Han, \etal, 2000), have not been carefully compared with detailed measurements of vortex pair motion in stratified environments and may also not accurately predict entrainment and detrainment rates.  It is clear that a more complete understanding of the physical mechanisms controlling the motion of vortex systems is needed.

The governing parameters for this flow are the Reynolds number $\Rey=\Gamma_0 / \nu$ , the vortex Froude number $\Fro = V_0/(N b_0)$, and the nondimensional turbulence intensity $\eta=(\epsilon b_0)^{1/3}/V_0$ of the background environment, in which $\Gamma_0$ is the vortex circulation strength, $\nu$ is the kinematic viscosity, $\epsilon$ is the turbulent dissipation rate, $b_0$ is the initial vortex separation distance and $V_0=\Gamma_0 /(2 \pi b_0)$ is the initial vertical speed of the vortex pair. The nondimensional turbulence intensity is the ratio of the turbulent velocity at the length scale of the vortex separation distance to the vortex descent speed. 

In this paper we describe two preliminary numerical experiments: the motion of a vortex pair for $\Rey = 10^5$, $\eta = 0.15$ and $\Fro = \infty$ and $4$. The values of these parameters are chosen such that the numerical simulations can be compared directly with aircraft measurements in an atmosphere with moderate ambient turbulence. These aircraft measurements are the only data we have access to for which the values of Re are near to that of full-scale submarine wakes.

The numerical method is the same large eddy scheme described in the previous section for computing the wake of a self-propelled body. However, for the vortex simulation we have modified the sub-grid-scale turbulence model in an initial attempt to deal with the effects of strong rotation, such as would be found in the vortex cores, on the small-scale turbulence. The flow is initialized with an approximation to a measured mean velocity field of a rolled up aircraft wing vortex, and a superimposed homogeneous turbulent velocity field in an unstratified background. A relaxation procedure is introduced, as described in the previous section, to allow the  turbulence field to establish a balance between production and dissipation before the calculation is begun.

\subsection{Problem Formulation}

\begin{figure}
   \begin{center}
     \includegraphics[width=\linewidth]{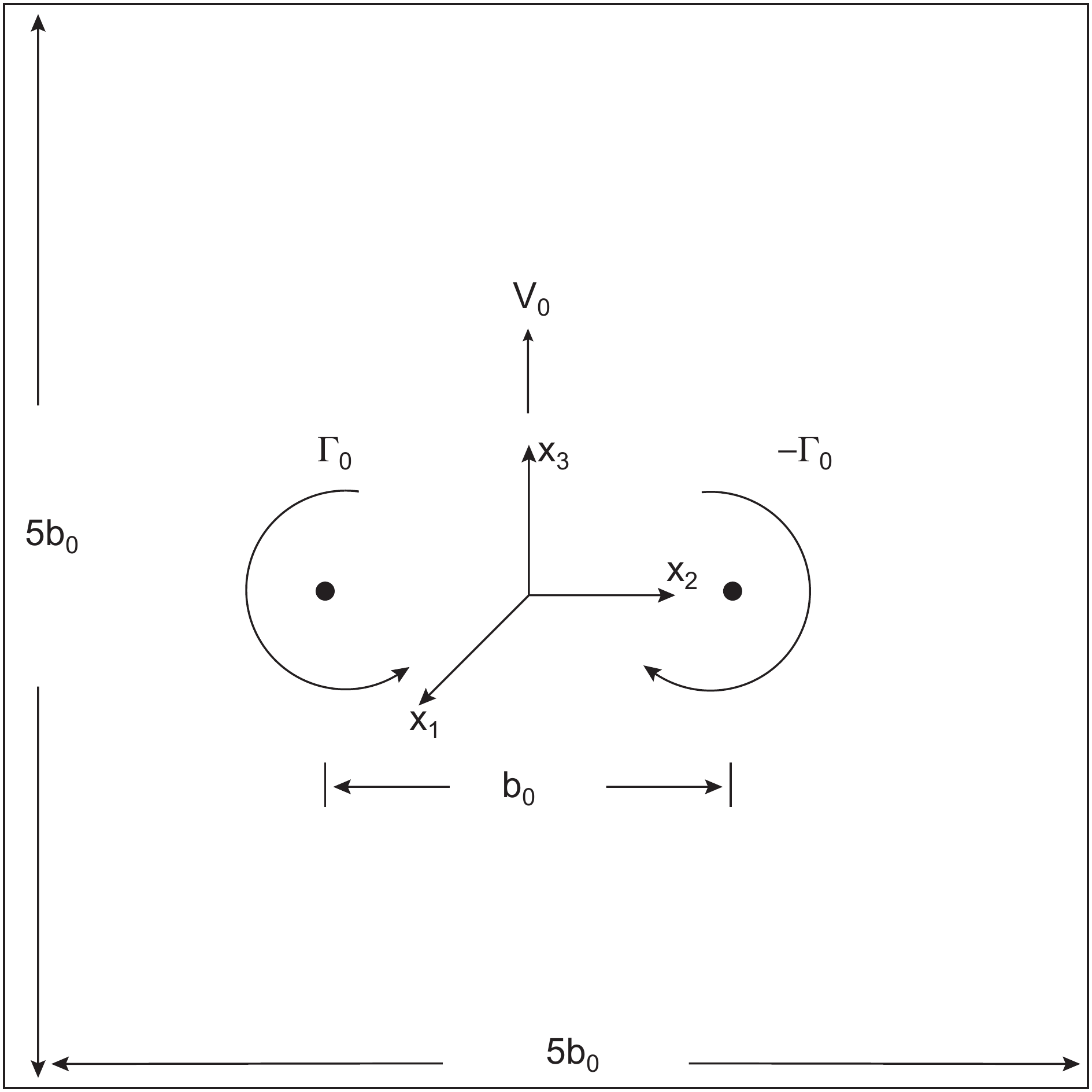}
   \end{center}
   \caption{A schematic diagram of a vortex pair in a stratified fluid, showing the coordinate system and relevant parameters, the vortex separation distance $b_0$, the vortex circulation strength $\Gamma_0$ and the self induced vertical velocity $V_0$.}  
   \label{vortex_schematic}
\end{figure}

A schematic drawing of the initial flow configuration for the vortex pair simulations is shown in Figure~\ref{vortex_schematic}, showing the locations of the centers of each vortex of the vortex pair. The distance between these centers is $b_0$ and the circulation strengths are $-\Gamma_0$ for the vortex on the left and $+\Gamma_0$ for the vortex on the right. The self-induced velocity of this vortex pair is directed upward and has magnitude $V_0$. The self-induced velocity is upwards since we are attempting to simulate a submarine that is slightly buoyant and so its control surfaces need to generate negative lift to keep the ship in level cruise. Superimposed on this vortex pair flow field is a homogeneous field of turbulence, whose intensity and characteristic length scale will be described below.

\subsubsection{Large eddy formulation of the Boussinesq equations} \label{sec:vortexLES}

We will assume that the fluid is incompressible and weakly stratified.  A large eddy approximation is invoked whereby the large-scale features of the flow are resolved and the small scales are modelled.  As in the previous section, let $\overline{u}_i$ denote the filtered three-dimensional velocity field as a function of space $x_i$ ($i = 1, 2, 3$) and time $t$.  The origin of the coordinate system is as shown in Figure~\ref{vortex_schematic}.  $x_1$ is along the axes of the two vortices, $x_2$ is transverse to these two axes, and $x_3$ is positive upward.  The length and velocity scales of the flow are respectively normalized by the initial vortex separation distance $b_0$ and the initial self-induced speed of the vortex pair $V_0$.

The numerical model used for this study is the same as that described in the previous {\it Large eddy formulation of the Boussinesq equations} subsection, except that the 
mixed model (Bardina, {\it et al.}, 1984) used to model the subgrid scale stress tensor has been modified in an initial attempt to represent the suppression of sub-grid-scale turbulence due to the strong rotation in the vortex cores. Since turbulence is suppressed in the cores, they tend to diffuse only very slowly. This modified mixed model was chosen as it is a first step towards implementing a dynamic SGS model, such as the one described by Germano, \etal.~(1991). For the modified mixed model,  the SGS stress tensor
\begin{eqnarray}
\tau_{ij}=\overline{u_j u_i} - \overline{u}_j \overline{u}_i \;\; ,
\end{eqnarray}
in which $u_i$ is the fluid velocity in the $i_{th}$ direction and the overbar symbol denotes spatial filtering, is represented as
\begin{eqnarray}
\tau_{ij} = \left( \overline{\overline{u}_j \overline{u}}_i
- \overline{\overline{u}}_j \overline{\overline{u}}_i \right)~~~~~~~~~~~~~~~~~~~~~~~~~~~~ \nonumber \\
-c_s \Delta^2 |\overline{S}_{ij}-\langle \overline{S}_{ij}\rangle| 
\left( \overline{S}_{ij}-\langle \overline{S}_{ij}\rangle \right) \;\; ,
\label{bardina}
\end{eqnarray}
where $S_{ij}$ is the strain tensor,
\begin{eqnarray}
\overline{S}_{ij} & = & \frac{1}{2} \left(
  \frac{\partial \overline{u}_i}{\partial x_j}
+\frac{\partial \overline{u}_j}{\partial x_i} \right) \;\; .
\end{eqnarray}

\noindent The first term, in parentheses, on the right hand side of (\ref{bardina}) is the similarity portion of the mixed model. It provides an accurate representation of the turbulent stresses. The remaining term on the right hand side is Smagorinsky portion portion of the mixed model. It represents dissipation and is formulated so that there is no turbulent dissipation when the strain tensor is well resolved. $c_s$ is the Smagorinsky coefficient and $\Delta$ is the width of the spatial filter, which we set equal to the grid spacing. In the simulations described here $c_s = 0.05$.

The computational domain is square in the cross plane with sides of length $5b_0$ and rectangular in the axial plane with the axial sides of length $1.5b_0$.  As in previous studies, the smaller length in the axial direction is chosen so as to suppress the Crow instability, Crow~(1970), (which has an axial wavelength of about $8.6b_0$) and to reduce computational costs. Periodic boundary conditions are imposed at all boundaries. In the calculations described in this paper, the grid resolution is $(64,256,256)$ in the $(x,y,z)$ directions.

\subsubsection{Initialization}

The initial flow field consists of the combination of a homogeneous turbulent velocity field and the velocity field associated with a vortex pair. The vorticity distribution within each vortex is an empirical fit to that measured in large-scale airplane wing vortices.

The background homogeneous turbulent velocity field is generated by prescribing an initially random distribution of Fourier modes and then integrating the equations of motion forward in time, holding the total energy constant by adjusting the amplitudes of all the Fourier modes appropriately at each time step, until the velocity field has reached a statistically steady state.  The resulting velocity field has a well developed inertial subrange and is nearly isotropic. After this steady state has been reached, the amplitudes are no longer adjusted and the turbulence is allowed to decay.  The density field is not perturbed initially. The strength of the turbulence field is measured by the parameter $\eta=\epsilon^{1/3}$, where $\epsilon$ is the nondimensional turbulent dissipation rate. This is a measure of the ratio of the turbulent velocity magnitude at the scale of the initial vortex separation to the self-induced speed of the vortex pair.

\begin{figure}
   \begin{center}
     \includegraphics[width=\linewidth]{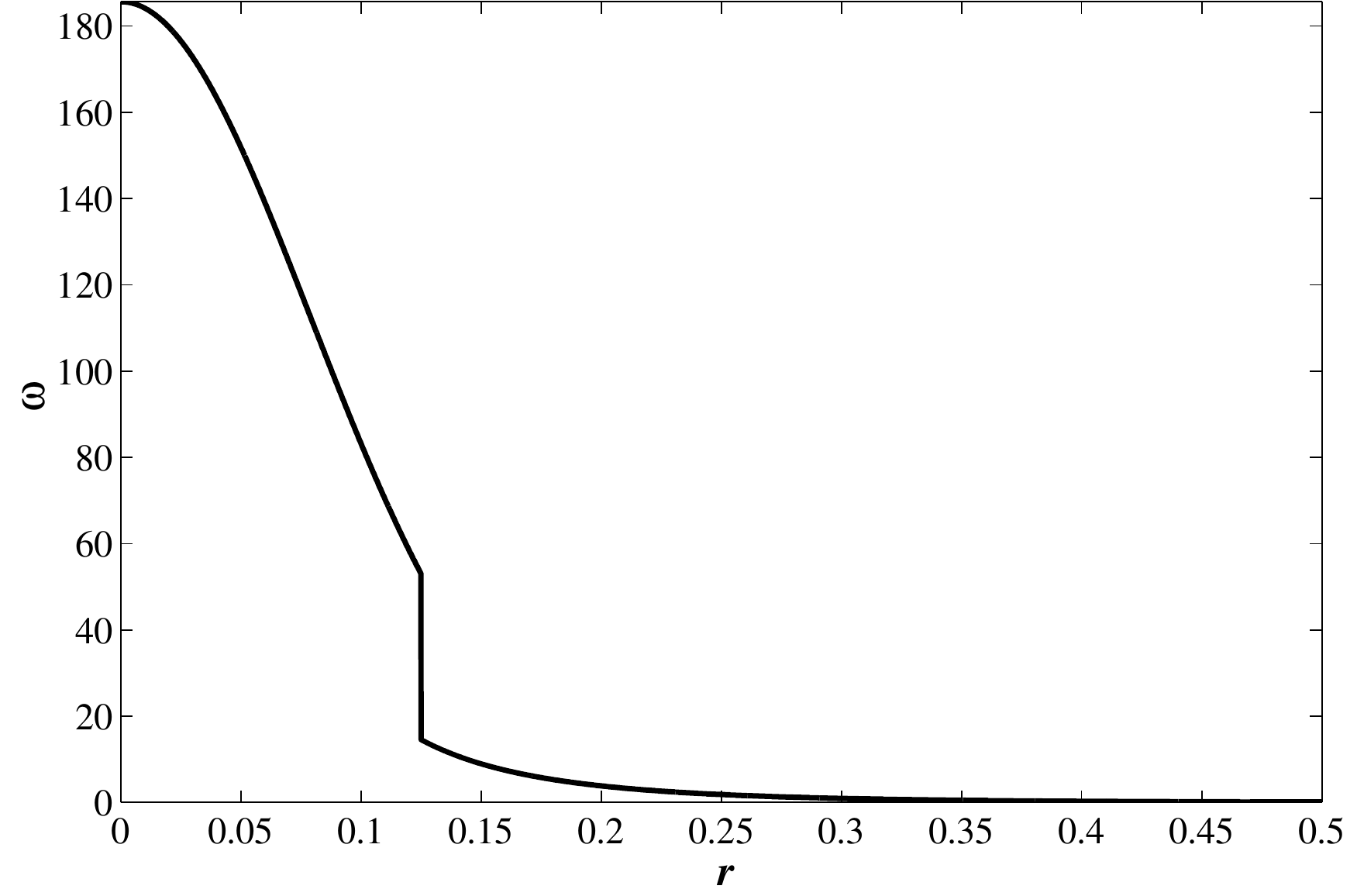}
   \end{center}
   \caption{The initial radial distribution of vorticity for each vortex of the vortex pair, normalized by their vorticity values at each vortex's center, plotted as a function of the radial distance, $r$, from the vortex center. The initial vorticity of each vortex is axisymmetric about the vortex center.}
   \label{vorticity}
\end{figure}

The initial velocity field associated with the vortex pair that we will use is an empirical fit to observed airplane wake vortices developed by Proctor~(1998). In this representation, the lateral components of the vorticity are zero and the axial component $\omega$ associated with each vortex is given by two functions, one for inside the vortex core, $r < r_c$, where $r$ is the radial distance from the vortex center and $r_c$ is the radial distance to the peak tangential velocity, and one for outside, $r > r_c$, the core. For $r < r_c$ 
\begin{eqnarray}
\omega(r) =  \frac{3.5076}{r_c^2} \left[1-exp[-10(r_c/B)^{3/4}]\right] \times \nonumber \\  exp[-1.2527(r/r_c)^2],
\end{eqnarray}
and for $r > r_c$
\begin{equation}
\omega(r) = \frac{7.5}{r B} (\frac{r}{B})^{-1/4} exp[-10(r/B)^{3/4}]
\end{equation}

\noindent in which $B = 4/\pi$ and to be consistent with airplane data $r_c = 0.125$. A plot of this vorticity distribution as a function of $r$ is shown in Figure~\ref{vorticity}.

The range of typical values of $\eta$ in the atmosphere, as reported by Han, \etal~(2000) is about 0.01 to 0.5. As a simulation representative of moderate turbulence, we chose to run the simulations described in this paper for $\eta = 0.15$.

The turbulent and vortex pair velocity fields are superimposed and adjusted so that the combined field is divergence free.

\subsection{Results}

Here we present preliminary results of the rate of ascent and the energetics of the vortex pair in non-stratified and stratified background environments. Of particular interest is the spatial distribution of the turbulent kinetic energy, which tells us where the most intense turbulence activity is going on and is a clue to where the most mixing and entrainment/detrainment into the vortex recirculating cell is going on.

\subsubsection{Vortex ascent rate}

\begin{figure}
   \begin{center}
     \includegraphics[width=\linewidth]{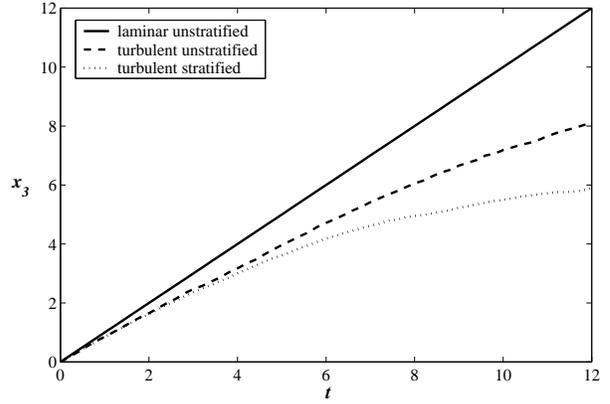}
   \end{center}
   \caption{The trajectory of the vortex pair for laminar unstratified conditions, turbulent unstratified and turbulent stratified.}
   \label{trajectory}
\end{figure}

The time series for the ascent of the vortex pair is shown in Figure~\ref{trajectory}. An ideal two-dimensional vortex pair in a non-turbulent and non-stratified fluid would ascend, in our nondimensional units, at a rate $Z = t$, in which $Z$ is the average vertical position of the pair.  As shown in Figure~\ref{trajectory}, the pair in a non-stratified but turbulent background ascends at a slightly slower rate than this. This appears to be due to the diffusion of vorticity by the turbulence. In a stratified fluid the vortex pair has to do work against buoyancy forces and this causes the pair to rise at an even slower rate. The rise rate is somewhat determined by how much fluid is entrained and detrained by the pair. If this entrainment and detrainment is large enough, then the pair will have a density that is always close to the density of the surrounding fluid, but if this entrainment is weak then the pair will have a density that is heavier than that of its surroundings and will stop ascending, much as what appears to be happening near $t = 12$ in Figure~\ref{trajectory} for the stratified case. 

The ascent rate for the vortex pair in a non-stratified fluid is consistent with the LES results of Han, \etal (2000) for the same strength of background turbulence. Han, \etal (2000) provide a semi-empirical formula for the prediction of the vortex ascent rate based on a diffusive decay of the circulation strength of the vortices. This formula seems to work well except when the background is strongly turbulent. By the end of the simulation period ($t = 12$) the center of the vortex pair has drifted to the left about $0.25b_0$ in the non-stratified case and about $0.5b_0$ in the stratified case.

\subsubsection{Energetics}

\begin{figure}
   \begin{center}
     \includegraphics[width=\linewidth]{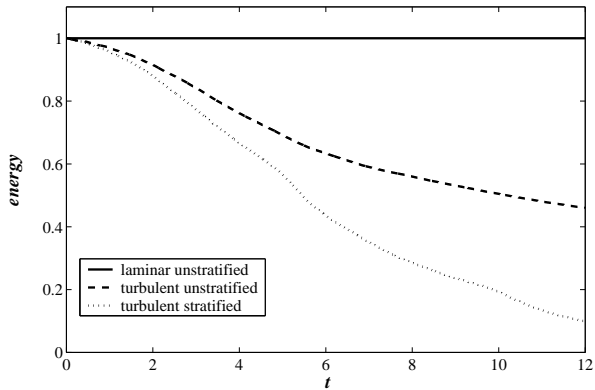}
   \end{center}
   \caption{The time series of the total kinetic energy of the mean flow.}
   \label{ke_mean}
\end{figure}

Time series of the kinetic energy of the mean flow, integrated over the whole computational domain, is plotted in Figure~\ref{ke_mean}. As can be seen, the mean energy decays steadily with time in both the non-stratified and stratified cases. The kinetic energy decays most rapidly in the stratified case since besides dissipation some of the kinetic energy is being converted to potential energy and some of the mean energy is being converted into internal wave fluctuations.

\begin{figure}
   \begin{center}
     \includegraphics[width=\linewidth]{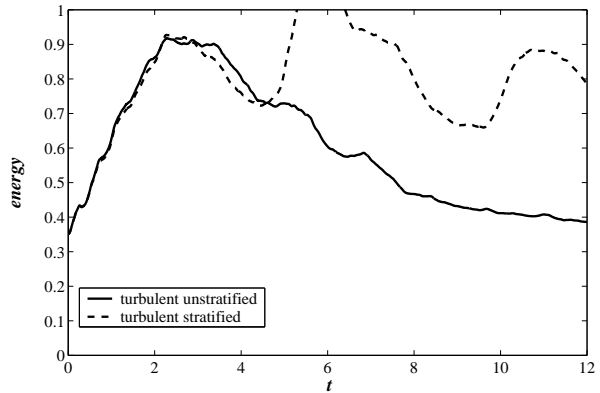}
   \end{center}
   \caption{The time series of the total kinetic energy of the turbulent flow.}
   \label{ke_fluct}
\end{figure}

Time series of the kinetic energy of the fluctuating motion in the whole computational domain is shown in Figure~\ref{ke_fluct}. In these time series, the fluctuating energy initially increases steadily as energy from the mean flow is converted to fluctuating energy. Up to $t \approx 1$, which is about a quarter of a buoyancy period, there is no difference between the non-stratified and stratified cases and there is not significant until $t = 2$ or $3$, which is about half a buoyancy period. At this time it appears that the turbulent energy saturates and begins to decline in the non-stratified case and to oscillate (with a slight decay) in the stratified case. The frequency of the oscillations in the stratified case is approximately equal to the buoyancy frequency, suggesting that a good fraction of the energy has gone into internal gravity waves.

\begin{figure*}
   \begin{center}
     \includegraphics[width=\linewidth]{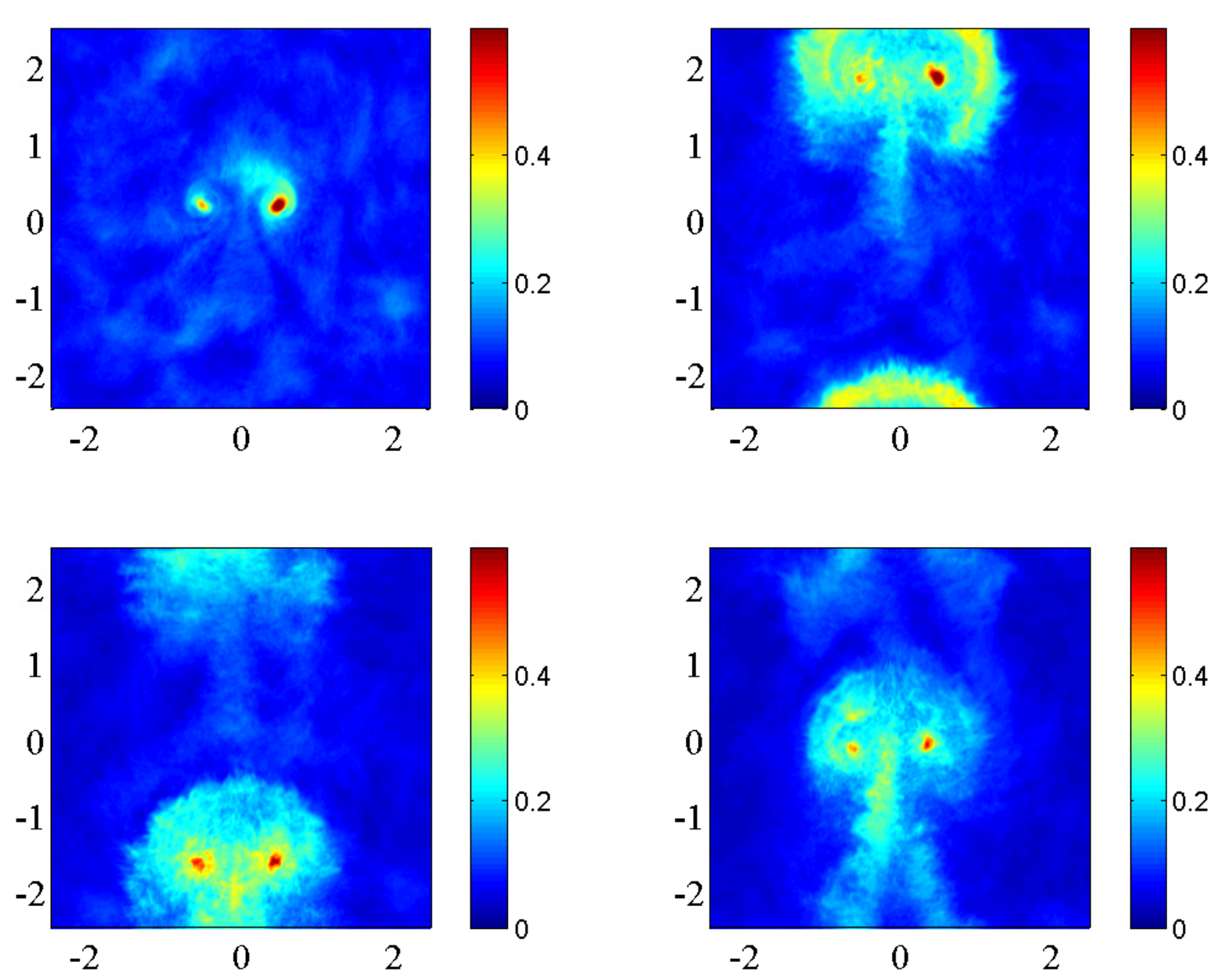}
   \end{center}
   \caption{Transverse plane $(y,z)$ contour plots of axially-averaged turbulent kinetic energy at (from left to right and top to bottom) $t = 0.2, 2.2, 4.2$ and $6.2$ for $\Rey = 10^5$ and $\Fro = \infty$.}
   \label{tene_unstratified}
\end{figure*}

\begin{figure*}
   \begin{center}
     \includegraphics[width=\linewidth]{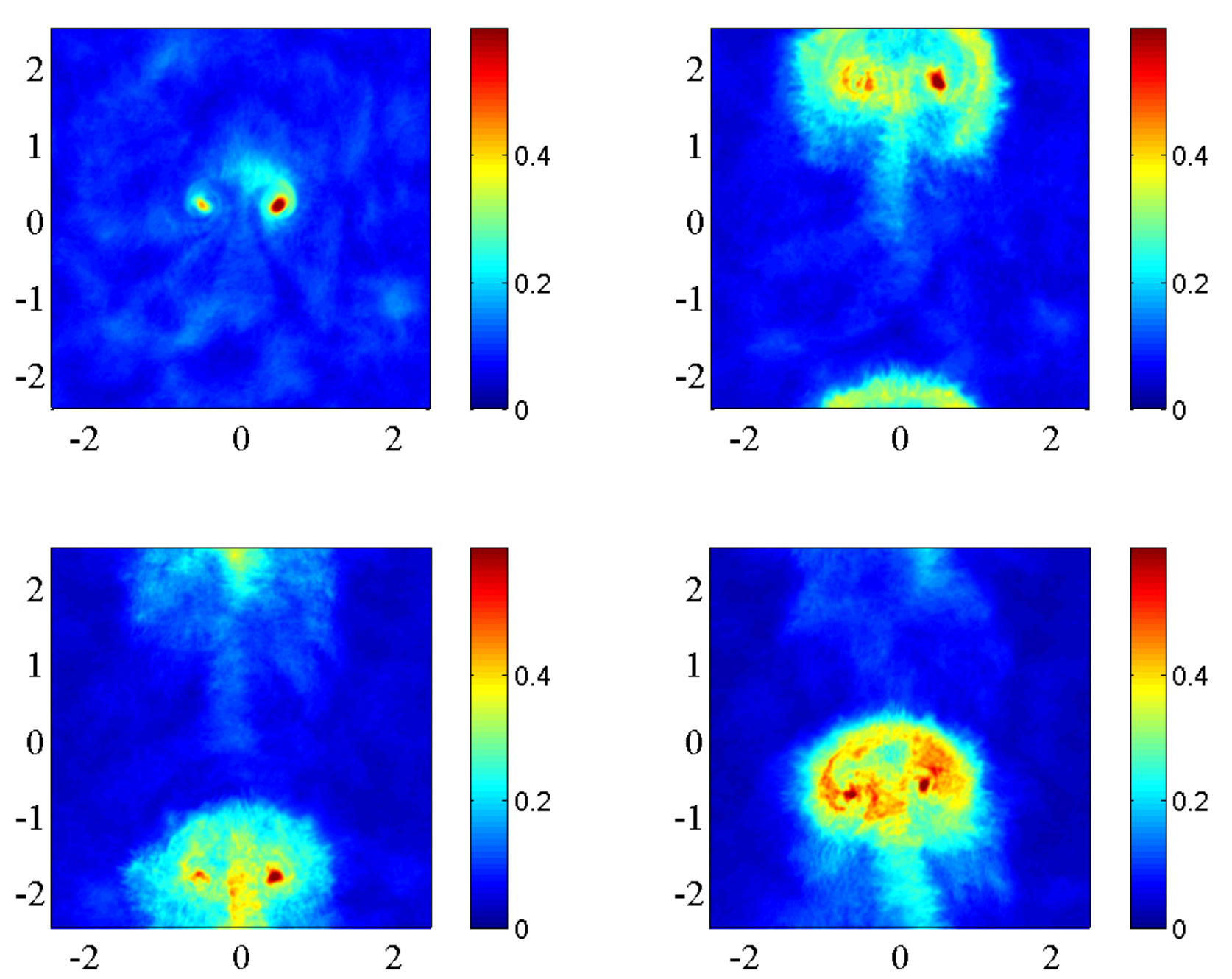}
   \end{center}
   \caption{Transverse plane $(y,z)$ contour plots of axially-averaged turbulent kinetic energy at (from left to right and top to bottom) $t = 0.2, 2.2, 4.2$ and $6.2$ for $\Rey = 10^5$ and $\Fro = 4$.}
   \label{tene_stratified}
\end{figure*}

The evolution of the spatial distribution of the turbulent kinetic energy is shown in Figures~\ref{tene_unstratified} and \ref{tene_stratified}.  Early in the simulation most of the turbulent kinetic energy is is the vortex cores in both the non-stratified and stratified cases. A little later, a significant amount of turbulent production has occurred at the top of the vortex pair recirculation zone and again this appears to be the same in both the non-stratified and stratified cases. After $t \approx 4$ the two cases start to deviate from each other substantially. In the non-stratified case, the turbulent kinetic energy remains concentrating in the cores as the energy decays steadily outside of the cores. In the stratified case, the turbulent kinetic energy actually increases at first in the region between the cores and eventually throughout the entire recirculation zone.

\section{Conclusions}

We have presented some preliminary results from using large eddy simulation to compute the late wake of a self-propelled body moving at constant speed through a non-stratified and a uniformly stratified fluid at $Re = 10^5$ as well as the motion of a pair of counter-rotating vortices in both non-stratified and stably stratified fluids.

An important aspect of the simulations is the use of a relaxation procedure to adjust the initial turbulence fields so that turbulent production and dissipation are in balance.  Our simulations intentionally have been initiated with different turbulent velocity and density fields than would be found in laboratory experiments.  We have done this so that the initial conditions would be nearly the same for stratified and non-stratified simulations as well as for simulations with different Reynolds numbers.  These similar initial conditions, which would be very difficult to obtain in laboratory experiments, have allowed us to study how the wake behaves without the complications of the initial conditions varying with the overall flow conditions.

We have observed that the pancake eddies form without being driven by similarly-sized eddies in the initial disturbance. In fact, these simulations produced pancake eddies even though the fluctuating component of the initialization began with random phase. We have seen that the eddies depend strongly on whether the body is towed or propeller driven: the propeller-driven body produces eddies that are smaller and more chaotic than the nearly periodic pancake eddies in the simulated wake of a towed body. This difference is consistent with the much more rapid decay of the mean axial velocity in the wake in the self-propelled case.

In the simulations of the motion of a vortex pair in a turbulent background flow in both non-stratified and stratified fluids we found that the ascent rate of the pair is strongly affected by the background turbulence and by the stratification, both effects lead to a decrease in the ascent rate of the pair. In the non-stratified case this is due to turbulent diffusion of the vorticity and in the the stratified case it is additionally due to the conversion of mean kinetic energy into potential energy and to the generation of internal gravity waves.

The distribution of turbulent kinetic energy in the vortex pair was studied and found to be quite different in the intermediate and late times of the simulations. The non-stratified simulation shows the turbulent kinetic energy mainly decaying everywhere except in the vortex cores. The stratified case shows that the turbulent kinetic energy actually increases in the recirculation zone at late times.

\vspace{1cm}

\section{Acknowledgements}
This research is supported by ONR under contract number N00014-01-C-0191, Dr.\ L.~Patrick Purtell program manager. This work was supported in part by a grant of computer time from the DOD High Performance Computing Modernization Program at the Naval Oceanographic Office Major Shared Resource Center. We thank Prof. G.~R.~Spedding at the University of Southern California and Prof. D.~D.~Stretch at the University of Natal for many helpful discussions.

\end{document}